\documentclass[]{aa}
\usepackage{graphicx}
\usepackage{txfonts}
\usepackage[longnamesfirst]{natbib}
\bibpunct{(}{)}{;}{a}{}{,}

\begin{document}

\title{Bimodal distribution of the autocorrelation function \\
    in gamma-ray bursts}

\author{Luis Borgonovo \inst{1}}

\offprints{L. Borgonovo, \email{luis@astro.su.se}} 

\institute{Stockholm Observatory, SE-10691 Stockholm, Sweden}

\titlerunning{Bimodal distribution of the ACF in GRBs}
\authorrunning{L.\ Borgonovo}

\date{Received: 24 Oct 2003 ; Accepted: 19 Jan 2004}


\abstract{
Autocorrelation functions (ACFs) are studied for a sample of 16 long
gamma-ray bursts (GRBs) with known redshift~$z$, that were observed by
the BATSE and Konus experiments. When corrected for cosmic time
dilation, the ACF shows a bimodal distribution. A {\it narrow} width
class (11 bursts) has at half-maximum a mean width $\tau'_o=1.6$~s with
a relative dispersion of $\sim 32$\%, while a {\it broad} width class
(5 bursts) has $\tau'_o=7.5$~s with a $\sim 4$\% dispersion. The
separation between the two mean values is highly significant
($>7\sigma$). This temporal property could be used on the large
existing database of GRBs with unknown redshift. The broad width set
shows a very good linear correlation between width at half-maximum and
$(1+z)$, with a correlation coefficient $R=0.995$ and a probability of
chance alignment $<0.0004$.  The potential application of this
correlation to cosmology studies is discussed, using it in combination
with recently proposed luminosity indicators.
}       
\maketitle




\section{Introduction} \label{intro} 

The knowledge of time scales and source distances are essential for
the physical understanding of astronomical phenomena. From the first
detections in 1969 by {\it Vela} satellites \citep{KSO73}, until the
launch of {\it BeppoSAX} in 1997, the distance scale of gamma-ray
bursts (GRBs) remained unsettled. This mission provided arc-minute
localization, leading to the discovery of a fading emission towards
lower energy bands, the so-called afterglows. Thereafter, burst
redshifts $z$ have been determined from spectroscopic analysis of the
afterglows or, in some cases, of their associated host galaxies,
proving that at least long-duration bursts are at cosmological
distances. So far, the redshift of no short-duration burst  has
been clearly determined \citep[although see][]{kul02}. In this paper
only the class of long GRBs will be considered (i.e., those with time
duration $> 2$ s).

Up to date, more than 30 burst redshifts have been spectroscopically
measured thanks to immediate follow-up observations. On the other
hand, there is a wealth of data from thousands of GRBs for which the
redshift is unknown. Most of these were detected by the Burst and
Transient Source Experiment (BATSE). Other important motivations to
find a redshift estimator based only on the gamma-ray prompt emission are
the lack of optical counterparts in some cases (the so-called {\it
dark} afterglows), and the difficulty of spectroscopically determine
redshifts beyond $z=5$ due to the Lyman alpha absorption.
In recent years, two empirical
relations have been discovered to estimate the luminosity distance
exclusively from the analysis of the gamma emission. One relates the
isotropic luminosity with the time lag between different energy channels
\citep{NMB00}, and the other with a variability parameter of the light
curve \citep{rei01}. Both luminosity correlations can be used to
derive luminosity distances and, assuming some specific cosmology, the
corresponding redshifts. Thus, from these correlations it has been
possible to estimate GRB luminosity functions and demographic
distributions \citep[see, e.g.,][]{Nor02,LFR02}. These first
estimations indicate that the GRB population may peak at redshift $z
\sim 10$, being then ideal probes of the early universe. However, the
luminosity functions derived in these works predict source counts
$N(>P)$, as a function of photon flux $P$, that differ significantly
from the observed one \citep{schmidt03}. Much better calibration of
these empirical relations is needed, and that will only be possible
with a much larger number of independent redshift determinations
covering a broader $z$ range.

Individual power density spectra (PDS) of GRB are in general very
diverse, but the longest bursts show power-law spectra extended over
two frequency decades. Shorter bursts also display this property by
averaging the PDSs of a large sample \citep{BSS98,BSS00}. This
underlying power-law behavior indicates the absence of any preferred
time scale.  The autocorrelation function (ACF) is the Fourier
transform of the PDS, therefore it contains in principle the same
information that can be visualized in a different way. The ACF gives
a measure of the correlation between different points in the light
curve that are separated by a given time lag.  Various efforts have
been made using these data analysis tools to find a temporal
characteristic that might correlate with the redshift and, e.g.,
\citet{CYC02} have found a weak correlation between the power-law index
of the PDS and~$z$. See also \citet{att03} for a proposed redshift 
indicator.

In this paper it will be shown that the ACF can be used to define
characteristic times that strongly correlate with the redshift. In
\S~\ref{methods} the data selection and the use of the ACF are
described. Next in \S~\ref{results} it is shown that the ACF corrected
for time dilation effects has a bimodal distribution, and that this
property could be used to construct an empirical relation to estimate
$z$. Finally, the results and their possible applications are
discussed in \S~\ref{discussion}.





\begin{figure*}
\centering
\includegraphics[width=17cm]{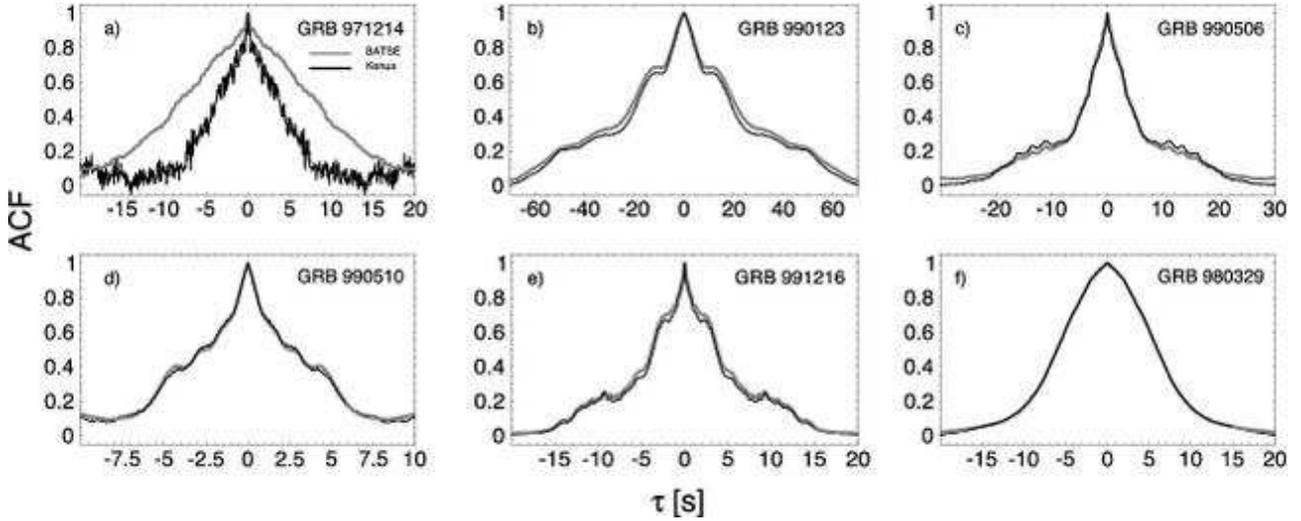}
\caption{Comparison of the ACFs of 6 GRBs obtained using data from two
different experiments. {\it Solid lines:} Konus 64 ms data in the
50--200 keV energy band; {\it gray lines:} BATSE 64 ms data in the
55--320 keV energy range. There is sufficiently good agreement for the
bright bursts, when the results are not very sensitive to the
background estimation. GRB 971214 is considerably dimmer than the
others (see the text for discussion).
\label{combo}}
\end{figure*}

\section{Data and methods} \label{methods}  

This work is mainly based on data taken by BATSE on board the {\it
Compton Gamma-Ray Observatory} \citep[CGRO;][]{Fish89}. It
consisted of eight modules placed on each corner of the satellite,
giving full sky coverage.  Each module had two types of detectors: the
Large Area Detector (LAD) and the Spectroscopy Detector (SD). The
former had a larger collecting area and from it the {\it CGRO} Science
Support Center (GROSSC) provided the so-called concatenated 64 ms
burst data, which is a concatenation of the three standard BATSE data
types DISCLA, PREB, and DISCSC. All three data types have four energy
channels (approximately 25--55, 55--110, 110--320, and $>320$
keV). The DISCLA data is a continuous stream of 1.024 s and the PREB
data covers the 2.048 s prior to the trigger time at 64 ms resolution,
both types obtained from the 8 LADs. They have been scaled to overlap
the DISCSC 64 ms burst data, that was gathered by the triggered LADs
(usually the four closer to the line of sight). This combined data
format was used when available, since the concatenated pre-burst data
allows a better estimation of the background.

All BATSE bursts with known redshift $z$ were considered for study
\footnote{See J. Greiner web page for a compilation of all GRBs with
known redshift at http://www.mpe.mpg.de/$\sim$jcg/grbgen.html}. In
some cases, like GRB 980326 and GRB 980613, the data are incomplete or
were not recorded. For burst GRB 970828 the DISCSC data are incomplete;
but it was possible to derive data with the same characteristics from
the MER data type, binning up the 16 energy channel into 4 DISCSC-like
energy channels. For  GRB 000131 the given DISCSC data are unevenly
sampled and it had to be uniformly binned into 1.024 s time
resolution. The BATSE sample total 11 cases.

To improve statistics, it was also considered GRB data that are
publicly available from other experiments. The Konus mission
\citep{Apt95} publishes GRB light curves of 64 ms resolution
within an energy band of 50--200 keV. At the time of this publication,
there were 25 Konus bursts with known redshift. But the collecting
area of this experiment is about 20 times smaller than the one on
BATSE and consequently, in most cases, the signal is too weak for the
purposes of this analysis.  A total of 5 bursts were selected for this
study (read further discussion below).

The INTEGRAL mission (launched in October 2002) makes public all count
time histories of the bursts detected by the anti-coincidence shield
of its gamma-ray spectrometer (SPI-ACS). They have a time resolution
of 50 ms and a non-sharp lower energy threshold at about 80 keV
\citep{KAL}. So far, the only detected burst with known $z$ is GRB
030329, and it was also detected by Konus. Therefore, these data were
used here mainly for comparative purposes.

The autocorrelation function of GRBs was first studied by
\citet{LRW93} and later on by, e.g., \citet{Fen95} and \citet{BSS00}. 
Following their notation, from a uniformly sampled count
history with $\Delta T$ time resolution and $N$ time bins, let $m_i$
be the total observed counts at bin $i$. Also let $b_i$ be the
corresponding background level and $c_i = m_i - b_i$ the net
counts. The discrete ACF as a function of the time lag $\tau = k
\Delta T$ is
\begin{equation}
A(\tau) = \sum_{i=0}^{N-1} \frac{c_{i} c_{i+k}}{A_o},\;\;\;\; k=1,
\ldots , N-1
\label{acf}
\end{equation}
and $A(0)=1$ for $k=0$. Here the periodic boundary conditions
($c_i=c_{i+N}$) are assumed. The normalization constant $A_o$ is
defined as
\begin{equation}
A_o = \sum_{i=0}^{N-1} c_{i}^2 - m_i .
\label{A0}
\end{equation}
The normalization makes the ACF of each burst fluence independent. The term
$m_i$ in Eq.~\ref{A0} subtracts the contribution of the
uncorrelated noise assuming that it follows Poisson statistics. 
This is not the case for the pre-trigger data of the concatenated BATSE
data type and they were excluded after the background fitting. Their
contribution to the ACF is negligible but they would affect the
estimation of $A_o$. This should also be considered when doing power
density spectral analysis. 
For practical reasons, the actual calculation of Eq.~\ref{acf}
was done using a Fast Fourier Transform (FFT) routine, i.e., squaring
the absolute value of the transformed gives the power density
spectrum, and taking the inverse transform of it gives the ACF. Zero
padding of the time series was used to avoid the artifacts produced by the
periodic boundary condition.

\begin{figure}
\resizebox{\hsize}{!}{\includegraphics{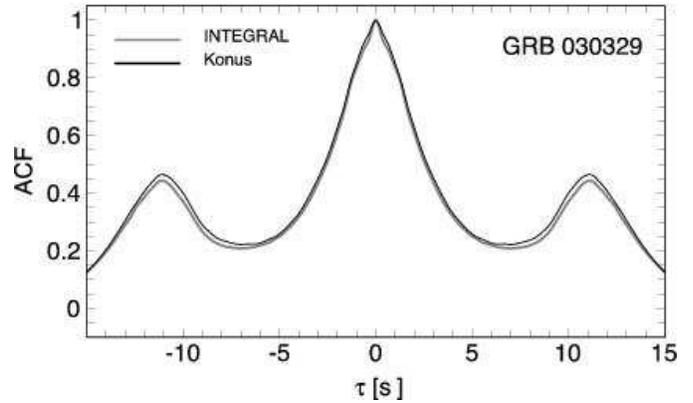}}
\caption{Two ACF functions of GRB 030329. {\it Solid
line:} Konus 64 ms data in the 50--200 keV energy band; {\it gray
line:} INTEGRAL 50 ms data with a soft low energy cut-off at $\sim
80$ keV. The high secondary peak of the ACF reflects the two bright
pulse structure of the light curve.
\label{IvsK}}
\end{figure}

The background estimation was done by fitting with up to second order
polynomial the pre- and post-burst data, that was judged by visual
inspection to be inactive. This is particularly critical for weak
bursts. Unfortunately, the Konus GRB light curves that are publicly
available have a fixed duration of 100 s, with no pre-burst data, and
sometimes not even post-burst data. Only a few cases are sufficiently
bright and have long post-burst data to allow a reliable estimation of
the ACF. The problem was studied using numerical simulations and it
became clear that for most of the Konus set the systematic errors
introduced by the background estimation are the main source of
uncertainty. Figure~\ref{combo} shows comparisons of the ACFs of
bursts for which there are data from both the Konus and BATSE
experiments. As reported by \citet{Fen95}, the ACF of GRBs narrows at
higher energies. The best match was obtained using the sum of the
BATSE energy channels 2 and 3, covering a similar energy range that
the corresponding Konus data. Note that the agreement will depend
mainly on having a similar lower-end energy limit, since there are
more counts at lower energies and the ACF is a quadratic function of
the number of counts.  The Konus weak case GRB 971214 illustrates how
a poor estimation of the background affects the ACF calculation. On
the other hand in the strong case GRB 990123, even with a short
post-burst data tail to fit the background, the difference between the
ACFs is acceptable for the purposes of this work. Guided by this
comparison, the selection criteria for the Konus cases were set,
requiring peak count rates larger than 3000 counts~${\rm s}^{-1}$ and
post-burst data.  These criteria are met by all bursts shown in
Fig.~\ref{combo} except the first, and by 5 other cases not observed
by BATSE that were then added to the sample. Among these last cases
is the bright GRB 030329 that was also observed by INTEGRAL, and
Fig.~\ref{IvsK} shows the good agreement between the ACFs derived
using the two different instruments data. Table~\ref{tsample}
summarizes in its four first columns the adopted sample of GRBs, the
instrument source, the estimated redshift $z$, and the corresponding
reference.

\begin{table}[t]
\caption{Sample of 16 GRBs with known redshift. The 6 columns give the
name of the GRB, the instrument, the measured redshift~$z$, the
corresponding reference, the ACF half-width at half-maximum
$\tau_{o}$, and the width corrected for time dilation $\tau'_{o}$.} 
\label{tsample} $$
\begin{array}{llllll}
\hline
\hline
\noalign{\smallskip}
\mathrm{GRB} & 
\mathrm{Instrument} &
z &
\mathrm{Ref.} & 
\tau_{o}(s) & 
\tau'_{o}(s) \\
\noalign{\smallskip}
\hline 
\noalign{\smallskip}
970508 & \mathrm{ BATSE }      & 0.835  & 1  & 2.70 &  1.47 \\
970828 & \mathrm{ BATSE}       & 0.9578     & 2  &    15.33 &  7.83 \\
971214 & \mathrm{ BATSE/Konus} & 3.418  & 3  & 8.02 &  1.81 \\
980329 & \mathrm{ BATSE/Konus} & 3\pm1  & 4  & 5.96 &  1.49 \\
980425 & \mathrm{ BATSE}       & 0.0085     & 5  & 7.62 &  7.56 \\
980703 & \mathrm{ BATSE}       & 0.966  & 6  & 14.15 &  7.19 \\
990123 & \mathrm{ BATSE/Konus} & 1.600  & 7  & 19.81 &  7.62 \\
990506 & \mathrm{ BATSE/Konus} & 1.3066     & 8  & 3.83 &  1.66 \\
990510 & \mathrm{ BATSE/Konus} & 1.619  & 9  & 2.54 &  0.97 \\
991208 & \mathrm{ Konus }      & 0.7055     & 10 & 3.67 &  2.15 \\
991216 & \mathrm{ BATSE/Konus} & 1.02   & 11 & 3.80 &  1.88 \\
000131 & \mathrm{ BATSE}       & 4.500  & 12 & 5.77 &  1.05 \\
000210 & \mathrm{ Konus }      & 0.846 & 13 & 2.43 &  1.31 \\
010222 & \mathrm{ Konus }      & 1.477  & 14 & 3.68 &  1.48 \\
011121 & \mathrm{ Konus }      & 0.362 & 15 & 9.98 &  7.32 \\
030329 & \mathrm{ Konus/INTEGRAL}  & 0.1685  & 16 & 2.56 &  2.19 \\
\noalign{\smallskip}
\hline
\end{array}
$$
\begin{list}{}{}
\item[1] \citet{mrm97}.
\item[2] \citet{djo01}.
\item[3] \citet{kul98}.
\item[4] \citet{lcr99}.
\item[5] \citet{tin98}.
\item[6] \citet{djo98}.
\item[7] \citet{kul99}.
\item[8] \citet{bloom03}.
\item[9] \citet{bea99}.
\item[10] \citet{dod99}.
\item[11] \citet{vea99}.
\item[12] \citet{And00}.
\item[13] \citet{pir02}.
\item[14] \citet{jha01}.
\item[15] \citet{gar03}.
\item[16] \citet{grei03}.
\end{list}
\end{table}


\section{Results}  \label{results}

\begin{figure*}
\centering
\includegraphics[width=17cm]{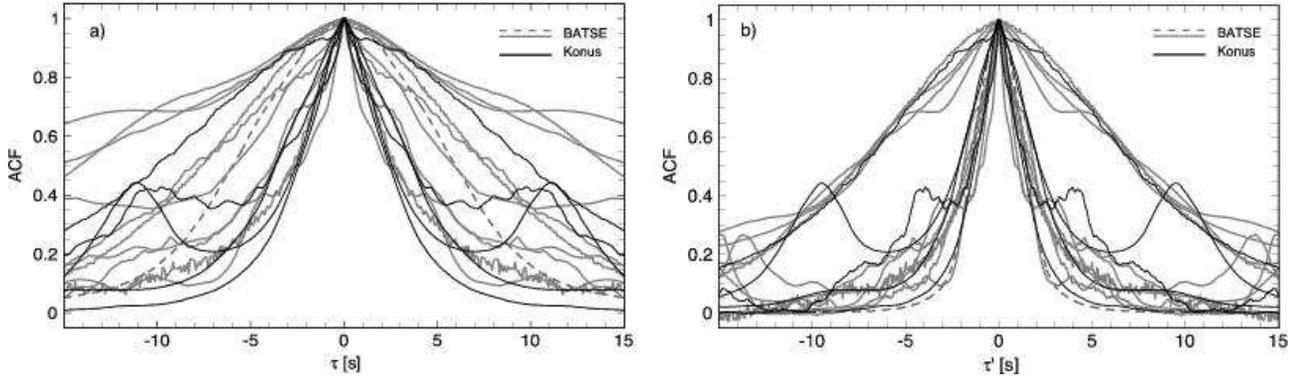}
\caption{{\bf a)} Autocorrelation functions of the 16 bursts
sample. BATSE and Konus cases are shown in gray and solid lines
respectively.  {\bf b)} Local ACFs, where time dilation due to cosmic
effect has been corrected, being $\tau' = \tau/(1+z)$. GRB 980329 is
shown with dashed lines assuming a redshift $z=3$.
\label{ACFs}}
\end{figure*}

In Fig.~\ref{ACFs}$a$ the ACFs of all the GRB sample are shown. One
can see that at different heights the width of the ACFs has a fairly
uniform distribution, with the half-width $\tau_o$ ranging, e.g., at
half-maximum between 2.5--20 s. Figure~\ref{ACFs}$b$ shows the {\it
local} autocorrelation function $A(\tau')$, where the cosmic time
dilation has been removed, and $\tau' = \tau/(1+z)$ is the corrected
time lag. The width of the different ACFs shows now, particularly around
the half-maximum level, a bimodal distribution with a clear gap
between two sets. A {\it broad} width set of 5 bursts and a {\it narrow}
width set of 11 bursts. The redshift of GRB 980329 is only known to be
in the range $z=$ 2.0--3.9 \citep{lcr99}. Thus, for
Fig.~\ref{ACFs}$b$ an average value $z=3$ was chosen, but in any
case for the given $z$ range its ACF will lie within the other narrow
width bursts. This burst was used here to show the bimodality but will
be excluded from the following calculations.
For the 11 BATSE bursts the local ACF distribution was analyzed at
different energy channels. Although narrower at larger energies, the
ACF shows the same clear bimodal distribution in all channels.   


\begin{figure}
\resizebox{\hsize}{!}{\includegraphics{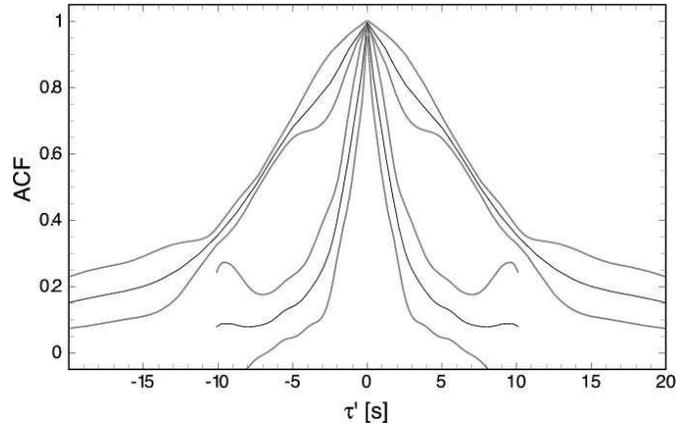}}
\caption{Mean value of the local ACF for narrow and broad width bursts
(solid lines). The $1 \sigma$ region about the mean is also shown
(gray lines). At half-maximum $A(\tau')=0.5$, $\tau'=(1.6\pm0.5)$~s
and $\tau'=(7.5\pm0.3)$~s for narrow and broad widths respectively.
\label{averages}}
\end{figure}

To study the distribution of the local ACF, e.g., to estimate
statistical moments at different lags, one cannot simply add the time
series since now all of them have time bins of different sizes. To
overcome this problem, the logarithm of each discrete ACF was
approximated by a polynomial function $ f(\tau)\simeq \ln A(\tau)$. A high
degree polynomial was used (typically $\sim 12$) to match the data within
the range of the random fluctuations up to time lags of 10 and 30 s
for narrow and broad cases respectively. These ranges were chosen to
well cover the central peaks of the ACFs down to the 0.1 level. Using
these functions, the mean and the sample standard deviation $s$ were
calculated for the two sets. Since the sample size $n$ is small in
both cases, the standard deviation $\sigma$ was estimated as 
\begin{equation}
\sigma = s \, (1+\frac{1}{\sqrt{2 (n-1)}}),
\end{equation}
\noindent using a correction 
for low number statistics assuming normal
distributions. Figure~\ref{averages} shows the mean curves and the
$\pm 1 \sigma$ region around them. At half-maximum $A(\tau'_o)=0.5$,
the half-width is $\tau'_o= (1.6\pm0.5)$~s and $\tau'_o=
(7.5\pm0.3)$~s for narrow and broad width sets respectively. Hence, at
half-maximum the distributions do not overlap even at the $7 \sigma$
level.  The probability of such separation into two sets by chance,
having an underlying unimodal distribution, was estimated numerically.
The overall distribution was assumed uniform in a given range (most
favorable case). Considering the same total number of points (15), and
asking conservatively the size of the smallest set to be $\ge 4$, the
probability $p$ of obtaining two sets with a difference between their
means $> 7 \sigma$ is $p < 6 \times 10^{-7}$. Therefore, the
alternative hypothesis of having a bimodal distribution can be
accepted at a highly significant confidence level.

\begin{figure}
\resizebox{\hsize}{!}{\includegraphics{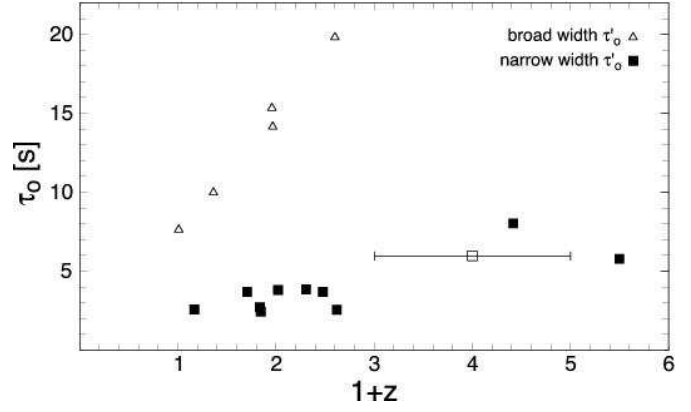}}
\caption{Correlation between the width at half-maximum  $\tau_o$ and the
dilation factor $1+z$. Bursts are classified into narrow and broad width
$\tau'_{o}$ cases, as seen by a local observer, and they are marked by
squares and triangles respectively. GRB 980329 is indicated with a
hollow square; its redshift is only known to be within the shown $z$
range. A linear fit of the form $f(x)=a x$ gives $a_n=(1.42\pm0.15)$~s
and $a_b=(7.54\pm0.10)$~s for narrow and broad sets. The slopes are
approximately equal to the average $\tau'_o$ for each set. Note that
since the slope ratio is 5.3, for a given $\tau_o$ each correlation
predicts very different $1+z$ values.
\label{corr}}
\end{figure}

The small range of the broad width distribution around the
half-maximum level is particularly interesting, because it represents
a relative dispersion of $4\%$, while the relative dispersion of the
narrow width is $32\%$. This means that if we had a way to know {\it a
priori} at what width class a burst belongs, we would be able to
estimate with the same corresponding uncertainties the time dilation
factor $1+z$, and therefore the redshift in practice only when $z
\gtrsim 1$.  The width $\tau_o$ was calculated fitting the logarithm
of the ACF in the range $0.4 \le A(\tau)\le 0.6$ with a second degree
polynomial. In the last two columns of Table~\ref{tsample}, the
obtained values for $\tau_{o}$ and $\tau'_{o} =\tau_{o}/(1+z)$ are
listed. Figure~\ref{corr} shows $\tau_{o}$ versus $1+z$ for both width
classes. As expected, the broad width set shows a very good
correlation, with a linear correlation coefficient $R=0.995$ and a
probability of chance alignment $p<0.0004$. The corresponding values
for the narrow width set are $R=0.809$ and $p<0.005$ respectively.
Notice that GRB980425, which has been associated with SN 1998bw,
belongs to the broad width set. This burst was considered an {\it
outlier} in the studies of the lag and variability luminosity
correlations when modeling the data with a single power-law, although
its inclusion supports the general trend in both cases.


\section{Discussion and conclusions}  \label{discussion}

The average PDS of bursts shows an overall power-law behavior,
indicative of a self-similar underlying process where there are no
preferred timescales. If this is the case, then the width of the ACF
is related with the low-frequency cut-off of the PDS, which is due to
the finite duration of the burst \citep{BSS00}. As it was mentioned in
\S~\ref{intro}, in principle the information given by the ACF and the PDS
is the same. In practice, since they express this information
differently, they are affected by noise and window effects in
different ways. It would be difficult to make a good estimation of the
low-frequency cut-off in the PDS due to the large statistical
fluctuations. On the other hand, the width at half-maximum of the ACF
gives a robust measure.

\begin{figure}
\resizebox{\hsize}{!}{\includegraphics{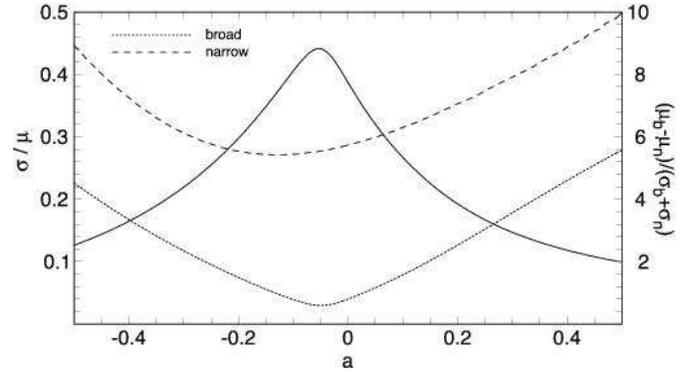}}
\caption{Variation of the dispersion of the ACF width with additional
redshift dependencies. The local values are calculated assuming $\tau'_{o}
=\tau_{o}/(1+z)^{1+a}$, where $a$ is an unknown correction index. The
left ordinate shows the relative dispersion of the broad ({\it dotted
line}) and narrow ({\it dashed line}) sets. Dispersion minima occur at
$a \ll 1$ in both cases. The right ordinate shows for the {\it solid
line} the difference between the mean values of each set ($\mu_b$,
$\mu_n$) in number of total standard deviations ($\sigma_b+\sigma_n$).
The {\it gap} between sets remains above $3\sigma$ over a large range
($-0.4<a<0.3$), indicating the robustness of the bimodality result
to additional corrections.
\label{index_a}}
\end{figure}

In this analysis the ACFs of bursts were only corrected for the cosmic
time dilation. However, since the detectors are sensitive over a
finite energy band, effects due to the shift in energy should also be
present. Studing a set of 45 bright long bursts, \citet{Fen95} found
that the full-width $W$ of the average ACF (at the $e^{0.5}$ level)
depends on the energy $E$ as $W(E) \propto E^{-0.4}$.  This narrowing
of the ACF should partially counteract the time stretching since for
large redshifts the energy window of the instrument will see photons
emitted at higher mean energies.  Furthermore, due to the trigger
threshold bursts detected at high redshifts are more luminous. There
are indications that the pulse width, and therefore the ACF width,
correlates with the luminosity \citep{LBP00}.
One should consider also that because of the energy shift bursts at
high redshifts are detected at earlier stages.
If the local ACF has an approximately constant width $\tau'_o$
(for each width class) these effects should produce a deviation from
linearity in Fig.~\ref{corr}.  Since no important deviation is
observed, the net combined effect must be small. To explore how
sensitive our results are to such effects we will assume now that the
local width of the ACF is given by $\tau'_{o} =\tau_{o}/(1+z)^{1+a}$,
where the index $a$ takes into account additional redshift
dependencies. Figure~\ref{index_a} shows the relative dispersion of
the width $\tau'_{o}$ for each set as a function of $a$. The
dispersion minima occur at small $a$ index values in both cases, with
$a_{\rm min}^{(b)}=-0.05\pm0.05$ and $a_{\rm min}^{(n)}=-0.13\pm0.23$
for the broad and narrow width sets respectively. The difference
between the mean values of each set versus $a$ is also shown in
Fig.~\ref{index_a}. It peaks at $a=-0.05$ with $8.8\sigma$, where now
$\sigma$ is the total standard deviation. Noteworthy, the {\it gap}
between sets remains larger than $3\sigma$ over a large range
($-0.4<a<0.3$), indicating how robust the bimodality result is to
any additional correction.


The practical use of the proposed empirical relations requires a
criterion to decide to what width class a burst belongs. For a width
$\tau_o \lesssim 7.5$ s the two classes do not overlap (see
Fig.~\ref{corr}). On the other hand, if extrapolations are valid,
$\tau_o \gg 20$ s implies for the narrow width class unrealistically
large $z$.  No burst spectral characteristic was found to correlate
with the width class. The same was true for the luminosity and total
energy release, but again larger samples should be studied to be
conclusive. 
The mentioned luminosity correlations could give a first $z$
estimation to determine the width class, and then it will be possible
using the ACF to obtain a second better and independent
estimation. Since the two classes are separated by more than a factor
5, the selection should not depend in practice on the assumed
cosmological parameters, and therefore the ACF width-redshift
correlation could be used in addition to constrain them.  

\begin{figure}
\resizebox{\hsize}{!}{\includegraphics{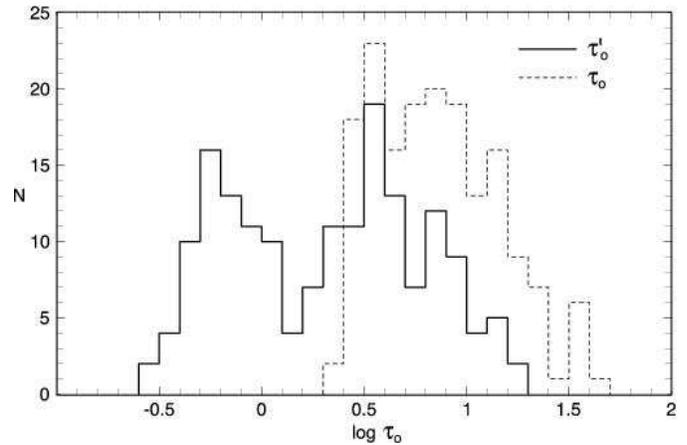}}
\caption{Distribution of the ACF width at half-maximum for a sample of
170 BATSE GRBs. {\it Dashed-line:} logarithmic histogram of the width
$\tau_{o}$ (in seconds). The distribution seems unimodal. {\it
Solid-line:} The width corrected for cosmic time dilation $\tau'_{o}$,
using estimated redshifts derived from the luminosity-variability
correlation \citep{FRR00}. The distribution appears now bimodal, with
median values approximately $0.7$ s and $4.0$ s for each subset.
\label{histo}}
\end{figure}

To investigate the combined use of these correlations, the ACFs of a
BATSE burst sample studied by \cite{FRR00} were calculated. They
estimated the redshifts of 220 GRBs using a power-law fit of the
luminosity $L$ versus the variability $V$ based on only 7
bursts. Therefore they obtained the best-fit parameters for their
model $L \propto V^{\alpha}$ with large uncertainties.  Furthermore,
the best-fit parameters vary considerably if GRB 980425 is included in
the calibration, and this burst was excluded in the derivation of the
published redshifts. Figure~\ref{histo} shows the logarithmic
distribution of the observed ACF widths $\tau_o$ derived for this
sample ({\it dashed-line}) and the distribution of the widths
$\tau'_{o}$ corrected for time dilation ({\it solid-line}). In
Fig.~\ref{histo} those cases where the iterative method to determine
$z$ diverged have been excluded, as well as bursts with observed ACF
widths $\tau_{o}<2.4$ s, but our conclusions do not depend on this
selection.  While the uncorrected distribution seems unimodal, the
corrected one appears bimodal, in qualitative agreement with the
results of \S~\ref{results}. The probability $p$ that a statistical
fluctuation could produce this feature was estimated.  Assuming an
underlying log-normal distribution, a conservative estimation gives
$p<0.02$. The $\tau'_{o}$ distribution is considerably broad, but that
is to be expected given the large spread found in the
luminosity-variability correlation. The median values for the two
width subsets are $\sim$$0.7$ s and $\sim$$4.0$ s, both approximately
a factor 2 smaller than the mean values of the distributions shown in
Fig.~\ref{averages}. However, this discrepancy can be accounted
considering the uncertainties.  Based on the analysis of 20 bursts
with known redshifts \cite{rei01} estimated
$\alpha=3.3^{+1.1}_{-0.9}$.  In particular, the exclusion of GRB
980425 will overestimate the exponent $\alpha$ and consequently the
redshifts, given smaller $\tau'_{o}$ corrected widths.

The ACF width-redshift correlations described inhere will need to be
confirmed by a larger statistical sample. Additionally, the lag and
variability luminosity correlations need to be known for a larger
redshift range to avoid uncertain extrapolations.  The close agreement
between the ACFs using data from the past mission BATSE and the
presently operating Konus and INTEGRAL will allowed to continue
improving the statistic of this work.  
In combination with the luminosity correlations we should be able to
construct a GRB-based Hubble diagram (i.e., a luminosity distance
versus $z$ plot) for high $z$, following a procedure similar to that
of \citet{sch03}. To obtain such diagram would have important
implications in cosmology studies. Ongoing efforts in this direction
will be presented in the near future.


\begin{acknowledgements}  
I wish to thank S.\ Larsson, C.~-I.\ Bj\"{o}rnsson, and F.\ Ryde for
useful comments and careful reading of the manuscript. This research
has made use of BATSE and Konus data obtained from the High Energy
Astrophysics Science Archive Research Center (HEASARC), provided by
NASA's Goddard Space Flight Center.
\end{acknowledgements}




\end{document}